\documentclass[aps,prl,twocolumn]{revtex4-1}
\usepackage{amssymb}
\usepackage{amsmath}
\usepackage{graphicx}

\begin{document}

\title{Twofold symmetry of proximity-induced superconductivity in Bi$_{2}$Te$_{3}$/Bi$_{2}$Sr$_{2}$CaCu$_{2}$O$_{8+\delta}$ heterostructures revealed by scanning tunneling microscopy}

\author{Siyuan Wan$^1$, Qiangqiang Gu$^1$, Huazhou Li$^1$, Huan Yang$^{1,*}$, J. Schneeloch$^2$, R. D. Zhong$^2$, G. D. Gu$^2$, and Hai-Hu Wen$^{1,\dag}$}

\affiliation{$^1$ National Laboratory of Solid State Microstructures and Department of Physics, Collaborative Innovation Center of Advanced Microstructures, Nanjing University, Nanjing 210093, China}

\affiliation{$^2$ Condensed Matter Physics and Materials Science Department, Brookhaven National Laboratory, Upton, New York 11973, USA}

\begin{abstract}
We observe proximity-induced superconductivity in the \textit{in situ} prepared heterostructures constructed by topological insulator Bi$_{2}$Te$_{3}$ thin films and high-temperature cuprate superconductors Bi$_{2}$Sr$_{2}$CaCu$_{2}$O$_{8+\delta}$. The superconducting gap maximum is about 7.6 meV on the surface of Bi$_{2}$Te$_{3}$ thin films with a thickness of two quintuple layers, and the gap value decreases with an increase in the film thickness. Moreover, the quasiparticle interference data show a clear evidence of a twofold symmetric superconducting gap with gap minima along one pair of the principal crystalline axes of Bi$_{2}$Te$_{3}$. This gap form is consistent with the $\Delta_{4y}$ notation of the topological superconductivity proposed in such systems. Our results provide fruitful information of the possible topological superconductivity induced by the proximity effect in high-temperature superconducting cuprates.
\end{abstract}

\maketitle

Topological superconductors (TSCs) with a pairing symmetry of odd parity host Majorana bound states which may play an important role in future applications of topological quantum computation \cite{SCzhangReview,FuReview}. A variety of approaches have been applied to achieve topological superconductivity after the initial theoretical predictions of topological nature in 2D $p+ip$-wave \cite{2DTSC} and 1D $p$-wave superconductors \cite{1DTSC}. One widely adopted method is to dope the topological insulators (TIs), for example, M$_{x}$Bi$_{2}$Se$_{3}$ (M = Cu, Sr, or Nb) \cite{CuBi2Se3c,CuBi2Se3d,SrBi2Se3,MBi2Se3}; the resultant superconductors have various properties related to the time-reversal-invariant topological superconducting states \cite{CuBi2Se3a,CuBiSeABS,CuBi2Se3b,CuBiSeARPES,STMCuBi2Se3a,STMCuBi2Se3b}. Theoretically, some iron-based superconductors are also predicted as possible candidates for TSCs \cite{FTStheory1,FTStheory2,FTStheory3,FTStheory4}, and experimentally Dirac-cone-type spin-helical surface states \cite{IBSARPES,FeSe1111,CaKFe4As4} as well as vortex cores with possible Majorana zero modes \cite{FeSeTe,FeSe1111,CaKFe4As4,FeTeSeHanaguri} are observed, which serve as possible evidence of topological superconductivity in these iron-based materials \cite{GuAPS}. Another approach to the realization of TSC is to construct TI/superconductor heterostructures, and the superconductivity in the TI layer induced by the proximity effect may be topologically nontrivial \cite{LFuProximity}. Such kind of superconductivity is successfully realized and proved in TI films grown on superconducting substrates of 2$H$-NbSe$_2$ \cite{Bi2Se3NbSe2,Bi2Te3NbSe2a,Bi2Te3NbSe2b} and FeTe$_{0.55}$Se$_{0.45}$ \cite{Bi2Te3FeSeTe,Bi2Te3FeSeTe2}. To date, cuprates have shown the highest superconducting critical temperature ($T_c$) record at atmospheric pressure, and the very large superconducting gap makes them a good candidate to induce proximity-induced superconductivity in the topological films made on top of them. According to theoretical predictions, the proximity effect may even be enhanced by the mismatch of the TI film and the cuprate Bi$_{2}$Sr$_{2}$CaCu$_{2}$O$_{8+\delta}$ (Bi2212) substrate \cite{Bi2Se3Bi2212theorya,Bi2Se3Bi2212theoryb}. By now, several attempts have been made on TI/Bi2212 heterostructures \cite{Bi2Se3Bi2212c,Bi2Se3Bi2212Zhou,Bi2Se3Bi2212a,Bi2Se3Bi2212b}. The gapped feature is observed on the Andreev reflection spectra measured in TI/Bi2212 junctions fabricated by the mechanical bonding technique \cite{Bi2Se3Bi2212c}, but the junction condition and the film thickness are not controllable. Afterward, a superconducting gap was observed in the Bi$_{2}$Se$_{3}$ film grown on Bi2212 substrate by angle-resolved photoemission spectroscopy (ARPES) measurements \cite{Bi2Se3Bi2212Zhou}, however, it was challenged by other works \cite{Bi2Se3Bi2212a,Bi2Se3Bi2212b}.

In TSCs with $D_{3d}$ crystalline symmetry, a twofold anisotropic superconducting gap seems to be a common feature. This twofold symmetric gap is observed in M$_{x}$Bi$_{2}$Se$_{3}$ materials by different kinds of measurements \cite{CuBi2Se3NMR,SrBi2Se3NMR,SrBi2Se3transport,CuBi2Se3transport}, and such gap breaks the threefold rotational symmetry of the crystal structures in these materials. The feature is explained as the spin-orbit interaction associated with the hexagonal warping effect which can induce a full superconducting gap with odd parity \cite{CuBi2Se3theory}. A twofold symmetric nodeless superconducting gap is also observed 1n 2 quintuple layer (QL) Bi$_{2}$Te$_{3}$/FeTe$_{0.55}$Se$_{0.45}$ heterostructures \cite{Bi2Te3FeSeTe}, and the obtained gap function is consistent with the $\Delta_{4y}$ gap notation predicted theoretically for TSCs \cite{CuBi2Se3theory} in the related system.

In this Rapid Communication, we report the successful deposition of TI Bi$_{2}$Te$_{3}$ thin films on Bi2212 substrates. We observe proximity-induced superconductivity on the surface of the heterostructures of Bi$_{2}$Te$_{3}$/Bi2212 by scanning tunneling microscopy/spectroscopy (STM/STS) measurements. A twofold symmetric superconducting gap is inferred from the twofold symmetric Fourier-transformed (FT-) quasiparticle interference (QPI) patterns at low in-gap energies. Our observations provide evidence of topological nature of proximity-induced superconductivity in Bi$_{2}$Te$_{3}$/Bi2212 heterostructures.

\begin{figure}
\includegraphics[width=8cm]{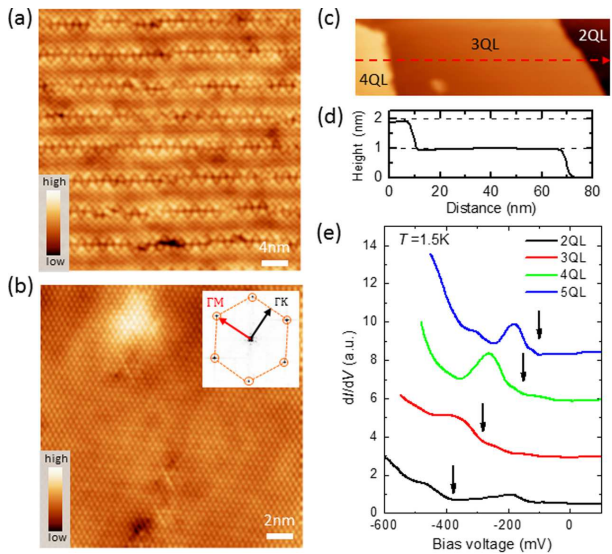}
\caption {Typical atomically-resolved topographic image of (a) the Bi2212 single crystal and (b) 2QL Bi$_{2}$Te$_{3}$ thin film grown on the top surface of Bi2212. The inset in (b) shows the FT image of (b). (c) Topography of Bi$_{2}$Te$_{3}$ film with different thicknesses. Set-point conditions: (a) $V_{set}=100$ mV, $I_{set}=50$ pA; (b) $V_{set}=50$ mV, $I_{set}=50$ pA; (c) $V_{set}=250$ mV and $I_{set}=50$ pA. (d) Spatial distribution of height measured along the arrowed line in (c). (e) A series of tunneling spectra measured on the heterostructures with different thicknesses of the Bi$_{2}$Te$_{3}$ layer. The arrows point out the kinks probably arising from the Dirac points of the surface states.
} \label{fig1}
\end{figure}

Optimally doped Bi2212 single crystals were grown by the floating-zone technique \cite{gugenda}. Figure~\ref{fig1}(a) shows the atomically resolved topography of a Bi2212 single crystal. The Bi$_{2}$Te$_{3}$ thin films are then successfully grown on the cleaved surface of Bi2212 by using molecular beam epitaxy technique \cite{MBE}. STM/STS measurements were carried out on \textit{in situ} prepared films. Detailed information on the film growth and STM/STS measurements are described in the Supplemental Material \cite{SI}. We show a typical atomically flat Bi$_{2}$Te$_{3}$ surface in Fig.~\ref{fig1}(b). The top atom layer of the film consists of Te atoms, and it has the hexagonal lattice structure with a lattice constant of about 4.3 \AA. The perfect hexagonal lattice can also be verified by the sharp and sixfold symmetric Bragg spots shown in the FT image in the inset of Fig.~\ref{fig1}(b). When we do the scanning in a relatively large area, we can observe some neighboring regions with different thicknesses. One example is shown in Fig.~\ref{fig1}(c). The height difference of about 1 nm across a step corresponds well to the height of a single QL of Bi$_{2}$Te$_{3}$. We regard these steps as the boundaries of the films with different thicknesses according to previous reports \cite{Bi2Te3NbSe2a,Bi2Te3NbSe2b,Bi2Te3FeSeTe,Bi2Te3FeSeTe2}. In order to determine the exact thicknesses of the film in different areas, we carry out tunneling spectrum measurements in a very wide energy range and show the spectra in Fig.~\ref{fig1}(e). One can see obvious kink features as marked by arrows; these kinks are supposed to be induced by the Dirac points on the topological surface states of Bi$_{2}$Te$_{3}$ \cite{Bi2Te3FeSeTe}. We can determine the thickness of the film in each area from the characteristic energy values of these kink features just as the operations in previous reports \cite{Bi2Te3NbSe2b,Bi2Te3FeSeTe}.

\begin{figure}
\includegraphics[width=8cm]{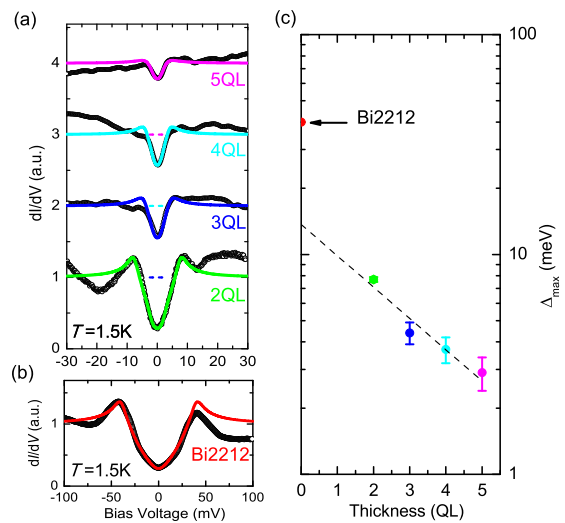}
\caption {A series of typical tunneling spectra (open circles) measured on (a) Bi$_{2}$Te$_{3}$/Bi2212 heterostructures and (b) the Bi2212 single crystal. The solid curves in (a) and (b) are the fitting results by using a one-gap Dynes model with (a) an anisotropic $s$-wave gap or (b) a $d$-wave gap. The short dashes near zero bias denote the zero differential conductance for the corresponding spectrum with the same color. (b) Semilog plot of the film-thickness-dependent superconducting gap maximum obtained from fittings. The error bars are determined in the fitting procedure by changing other fitting parameters.
} \label{fig3}
\end{figure}

Figure~\ref{fig3}(b) shows a typical tunneling spectrum measured on the Bi2212 single crystal after cleavage in an ultrahigh vacuum and before annealing or the film growth process; the finite zero-bias differential conductance may be due to the impurity scattering in the Bi2212 sample with a nodal gap \cite{Bi2212Gu,Bi2212i1,Bi2212i2,Bi2212i3}. Figure~\ref{fig3}(a) shows some typical tunneling spectra measured on the Bi$_{2}$Te$_{3}$/Bi2212 heterostructures, and these spectra have an obvious gapped feature. The tunneling spectra are roughly homogeneous in the films (see Figs.~S1 and S2 \cite{SI}), which indicates that the Bi$_{2}$Te$_{3}$ films are homogeneous with a high quality. With an increase of film thickness, the zero-bias conductance increases continuously, while the gap value decreases steadily. The latter is determined by the energy difference between the two coherence peaks. Because of the relatively large zero-bias differential conductance, the exact value of the superconducting gap should be obtained from the fitting procedures. Then we use the Dynes model \cite{Dynes} with a single gap to fit the measured tunneling spectra. For the spectrum measured on 2 QL heterostructures, the Dynes model with an isotropic $s$-wave gap cannot fit the spectrum well (see Supplemental Material \cite{SI} for details), hence the gap should be anisotropic in the heterostructures. From the fitting procedures, it is impossible to determine whether there are gap nodes on the gap function for the spectra measured on Bi$_{2}$Te$_{3}$/Bi2212 heterostructures. However, we find that the gap maximum $\Delta_{max}$ is almost independent of the gap function, e.g., $\Delta_{max}=7.6\pm0.2$ meV for the 2 QL film. In Fig.~\ref{fig3}(a), we show some typical fitting results for the spectra measured on the heterostructures by using an anisotropic $s$-wave gap, and the gap maximum values are shown in Fig.~\ref{fig3}(c). One can see that $\Delta_{max}$ decreases with an increase of the thickness of the Bi$_{2}$Te$_{3}$ layers following an exponential decay law approximately. Similar results were observed in the TI/2$H$-NbSe$_2$ heterostructures \cite{Bi2Se3NbSe2,Bi2Te3NbSe2a}. As shown in Fig.~\ref{fig3}(a), the linear extension value $\Delta_{max}(0\ \mathrm{QL}) =13.5$ meV is much smaller than the gap maximum 43 meV of Bi2212, which may be understood in the theoretical framework of proximity-induced superconductivity \cite{proximityeffect}.

\begin{figure}
\includegraphics[width=8cm]{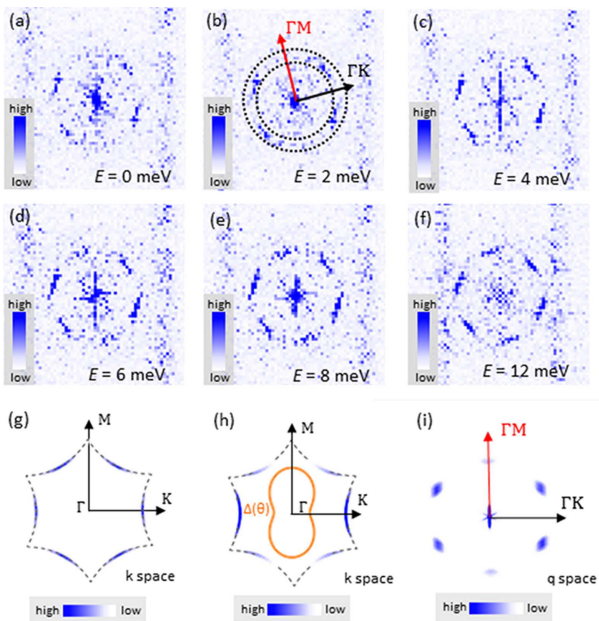}
\caption {(a)-(f) The FT-QPI patterns measured at different energies from 0 to $+12$ meV on the 2QL Bi$_{2}$Te$_{3}$/Bi2212 heterostructure ($T=1.5$ K). (g) Schematic figure of the surface state in the normal state and (h) that when the DOS is partially gapped in the presence of a twofold symmetric gap. The angular dependence of the anisotropic superconducting gap is also shown schematically by the orange solid line in (h). (i) Simulated FT-QPI pattern by doing self-correlation to (g) and considering the spin selection rules.
} \label{fig5}
\end{figure}

The superconducting gap anisotropy can be detected \cite{DavisLiFeAs,ZYDuNC} by the QPI measurements \cite{HoffmanReview}. In order to get detailed information on the superconducting gap in the heterostructures, we measure the QPI images at different energies (see Fig.~S5 \cite{SI}) on the 2QL Bi$_{2}$Te$_{3}$ films and show the corresponding FT-QPI patterns in Figs.~\ref{fig5}(a)-\ref{fig5}(f). In normal-state FT-QPI pattern shown in Fig.~\ref{fig5}(f), one can see six clear spots along the $\Gamma$M directions instead of a continuously distributed pattern from a complete Fermi surface, and the intensities of the six spots are almost the same. The FT-QPI results obtained on the 2QL Bi$_{2}$Te$_{3}$/Bi2212 heterostructures are similar to the results measured on Bi$_2$Te$_3$ films grown on Si substrates \cite{QPIBi2Te3film}, and the pattern can be interpreted as the contributions from the off-plane spin orientations \cite{Bi2Te3QPISCZhang,Bi2Te3QPIJPHu} when considering the hexagonal warping effect \cite{FuLWarping}. From our previous FT-QPI results measured on 2QL Bi$_2$Te$_3$/FeTe$_{0.55}$Se$_{0.45}$ heterostructures, there are no obvious scattering spots observed at zero bias \cite{Bi2Te3FeSeTe}, and we argue that the superconducting gap is nodeless for the proximity-induced superconductivity on the Bi$_{2}$Te$_{3}$ film. In contrast here on the 2QL Bi$_2$Te$_3$/Bi2212 heterostructure, one can see that four characteristic scattering spots appear even at zero bias from Fig.~\ref{fig5}(a) although the intensity of these spots is very weak when compared with that of spots measured in normal state [Fig.~\ref{fig5}(f)]. However, this does not mean that the gap should be nodal, and may suggest that the gap minimum is quite small.
In addition, the existence of the scattering spots at zero bias may be related to the finite density of states (DOS) at zero bias which is characterized by the gap filling effect appearing on the tunneling spectrum. Concerning the gap filling on the Bi$_{2}$Te$_{3}$ films deposited on Bi2212, there may be three possibilities, namely, i) proximity induced in-gap states from Bi2212; (ii) possible nodes in the induced superconducting gap of Bi$_{2}$Te$_{3}$; and(iii) topological surface states in Bi$_{2}$Te$_{3}$ films. About the first possibility, the gap filling effect already appears on the tunneling spectrum measured on Bi2212 [Fig.~\ref{fig3}(b)], and the DOS at zero bias in the 2QL film is even higher than that of Bi2212. Thus we believe that the gap filling in the 2QL film is unlikely coming from the substrate states by the proximity effect. The second possibility is that the induced topological superconducting gap may have nodes, which certainly yields some finite DOS with the presence of the impurities. The last possibility is more straightforward, that the topological surface state on the Bi$_{2}$Te$_{3}$ thin film may contribute finite DOS for itself since on the surface it is gapless although the superconducting gap opens in the bulk of Bi$_{2}$Te$_{3}$ film. At this moment, we cannot explicitly judge which possibility dominates here. Clearly this needs further investigations for the gap filling effect on Bi$_{2}$Te$_{3}$ films grown on Bi2212 substrates.

At low energies below 6 meV, one can see that a couple of scattering spots along one pair of $\Gamma$M directions have very weak intensities when compared with the spots in the same area on the normal-state FT-QPI pattern measured at $+12$ meV. The relatively weak intensities of the scattering spots along $\Gamma$M directions at low energies suggest the gap maximum in this direction. To further strengthen our argument, we try to simulate the FT-QPI pattern in the presence of a twofold anisotropic $s$-wave gap. First we adopt the Fermi surface of the system which has a sixfold symmetry as shown in Fig.~\ref{fig5}(g), and then we multiply the intensity of each $k$-point by a factor of $\sin^2\theta$ with $\theta$ the angle beginning from the vertical $\Gamma$M direction. The final angular dependent intensity is shown as the color plot of the outer contour in Fig.~\ref{fig5}(h). One can see clearly the twofold symmetric DOS distribution along the Fermi surface. However, we must emphasize that this serves only as a qualitative description. By doing self-correlation to Fig.~\ref{fig5}(g) and considering the spin selection rules \cite{Bi2Te3QPISCZhang,Bi2Te3QPIJPHu}, we obtain the simulated FT-QPI pattern and show it in Fig.~\ref{fig5}(i). A pair of scattering spots along one of the $\Gamma$M directions is very weak, which agrees well with the experimental data. Hence, we conclude that the gap maxima are along the $\Gamma$M directions and the gap minima are along one pair of the principal crystalline axes or the $\Gamma$K directions. It should be noted that the gap minima directions in the 2QL Bi$_2$Te$_3$/Bi2212 heterostructures are the same as those on the 2QL Bi$_2$Te$_3$/FeTe$_{0.55}$Se$_{0.45}$ heterostructures \cite{Bi2Te3FeSeTe}, although the characteristic scattering spots are along different directions when they are measured outside the gap energy (see Supplemental Material \cite{SI}, which includes Refs. \cite{S10,S11} for a detailed discussion).

\begin{figure}
\includegraphics[width=8cm]{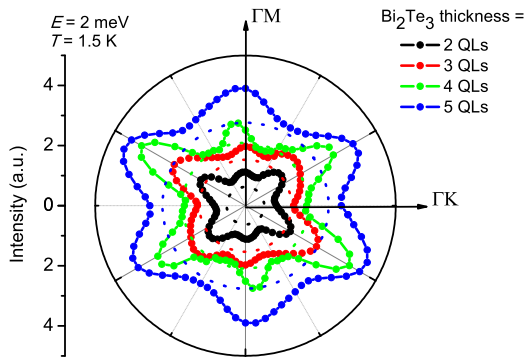}
\caption {Angle dependence of averaged intensity of the scattering spots measured on heterostructures with different thicknesses of Bi$_2$Te$_3$ but all at $E=+2$ meV. The averaged background intensities are also plotted as dotted lines by using the same color as the corresponding averaged curve in this figure. Both the experimental integrated curves and the background dotted lines are shifted for clarification.
} \label{fig6}
\end{figure}

In order to illustrate the QPI intensity variation more clearly, we calculate the angle dependence of the averaged FT-QPI intensity for the FT-QPI patterns measured on heterostructures with different thicknesses of Bi$_2$Te$_3$ (see Fig.~S8 \cite{SI}). As an example for the data measured at $+2$ mV in the 2QL film, we calculate the averaged FT-QPI intensity for each angle $\theta$ by averaging intensities of all the data points between the two circles as illustrated in Fig.~\ref{fig5}(b) and in the angle range of $\theta\pm 5$ degrees. Figure~\ref{fig6} shows the obtained angular dependent averaged intensity curves for the characteristic scattering spots at $E=+2$ meV on the heterostructures with different thicknesses of Bi$_{2}$Te$_{3}$. One can see very clear twofold symmetric intensity distributions on the heterostructures with Bi$_{2}$Te$_{3}$ thicknesses of less than 4 QLs, and the weaker scattering spots are all along one pair of $\Gamma$M directions. Even for the data measured on the 5QL heterostructure, some weak anisotropy can still be observed. The weakening of anisotropy on the heterostructures with thicker Bi$_{2}$Te$_{3}$ films is understandable since the differential conductance increases rapidly within the gap. Therefore, a twofold symmetric superconducting gap with gap maxima along one pair of $\Gamma$M directions can be observed in the Bi$_{2}$Te$_{3}$/Bi2212 heterostructures.

It is worth stressing that we observe proximity-induced superconductivity on the Bi$_2$Te$_3$/Bi2212 heterostructures from the tunneling spectra and QPI results. The superconducting gap has not been observed on the Bi$_{2}$Se$_{3}$/Bi2212 heterostructures in some ARPES measurements \cite{Bi2Se3Bi2212a,Bi2Se3Bi2212b}, and the authors argue that one possible reason for this is because of the very short coherence length of Bi2212 along the $c$-axis. In the current work, the gapped feature exists on the Bi$_2$Te$_3$/Bi2212 heterostructures with a thickness of more than 5 nm (5 QLs) which is much larger than the coherence length values of Bi2212 single crystals ($\xi_{ab}=0.38$ nm and $\xi_c=0.16$ nm in the zero-temperature limit) \cite{Bi2212cohlength}. It should be noted that for the proximity-induced superconductivity from a superconductor to a closely contacted metal, the superconducting effective range in the normal metal has no clear relationship to the coherence length of the superconductor \cite{proximityeffect}. For the proximity effect in heterostructures with copper oxide as the substrate, such as Bi2212 used here, because the superconductivity of the top layer in the cuprate is very sensitive to the annealing condition and the $c$-axis coherence length is very short, both can easily induce a degraded order parameter on the top layer. This may be the reason for different results coming out of different groups.

We observe twofold symmetric QPI patterns in 2QL Bi$_2$Te$_3$/Bi2212 heterostructures at small in-gap energies, which naturally suggests a twofold symmetric gap for the proximity-induced superconductivity in the Bi$_2$Te$_3$ films. One may argue that the twofold nature is related to the supermodulations of the Bi2212 substrate \cite{supermodulation} or the anisotropy of the FT image of the topography. In our point of view, this is unlikely and a detailed discussion is included in the Supplemental Material \cite{SI}. Although we can not judge whether the gap has nodes on the 2QL Bi$_2$Te$_3$/Bi2212 heterostructures, the gap minimum direction determined here allows us to conclude that the nodeless $\Delta_{4y}$ notation is a more possible gap structure in the present system. Since this gap notation is proposed theoretically for a topological superconductor \cite{CuBi2Se3theory}, our work provides extra evidence for the existence of topological superconductivity induced by proximity effect in the Bi$_2$Te$_3$/Bi2212 heterostructures.

To conclude, we successfully achieve proximity-induced superconductivity on \textit{in situ} grown Bi$_{2}$Te$_{3}$ films with different thicknesses on the cleaved surface of the high-$T_c$ cuprate Bi2212. The superconducting gap maximum on the 2QL Bi$_2$Te$_3$/Bi2212 heterostructures is as large as 7.6 meV, and the gap feature remains even when the topological insulator film is as thick as 5 QLs. The intensity of FT-QPI patterns show the twofold symmetric nature when measured at energies within the superconducting gap maximum, which suggests a twofold symmetry of the superconducting gap on the Bi$_{2}$Te$_{3}$ film. The orientation of the gap minimum is along one of the principal crystalline axes, which is consistent with the theoretically proposed $\Delta_{4y}$ notation. Our observations provide clear evidence of proximity-induced superconductivity possibly of a topological nature on these kinds of heterostructures consisting the TI and the high-$T_c$ cuprates.

\begin{acknowledgments}
We acknowledge J. F. Jia for sharing knowledge of the deposition technique for the Bi$_{2}$Te$_{3}$ thin film. The work was supported by the National Key R\&D Program of China (Grant No. 2016YFA0300401), the National Natural Science Foundation of China (Grant No. 11534005 and No.11974171), and the Strategic Priority Research Program of Chinese Academy of Sciences (Grant No. XDB25000000). The work at Brookhaven was supported by the Office of Basic Energy Sciences, U.S. Department of Energy (DOE) under Contract No. DE-SC0012704.
\end{acknowledgments}

$^*$ huanyang@nju.edu.cn

$^{\dag}$ hhwen@nju.edu.cn

\end{document}